\documentclass[journal,onecolumn]{IEEEtran} 

\usepackage[utf8]{inputenc}
\usepackage[T1]{fontenc}
\usepackage{amsmath,amssymb,amsfonts}
\usepackage{bm} 
\usepackage{siunitx} 
\usepackage{amsmath} 
\usepackage{empheq} 
\usepackage{xcolor} 
\usepackage{hyperref}

\newcommand{\herm}{H} 
\newcommand{\pr}{\partial}

\title{Derivation of CRB and Refined SINR Expressions for OTFS-RSMA LEO ISAC Systems}
\author{Bruno Felipe Costa and Taufik Abrão
\thanks{This document provides supplementary derivations for the paper "An Integrated OTFS-RSMA Framework for LEO Satellite ISAC: Modeling, Metrics, and Potential". The methodologies employed for handling practical impairments are grounded in established literature.} 
}
\date{\today}

\begin{document}

\maketitle

\begin{abstract}
This document provides detailed step-by-step derivations for the Cramér-Rao Bounds (CRB) for sensing parameters and the refined Signal-to-Interference-plus-Noise Ratio (SINR) expressions under imperfect Channel State Information (CSI) and imperfect Successive Interference Cancellation (SIC) for the Orthogonal Time Frequency Space (OTFS) Rate-Splitting Multiple Access (RSMA) framework presented in our main work "An Integrated OTFS-RSMA Framework for LEO Satellite ISAC: Modeling, Metrics, and Potential". These derivations support the analytical expressions and models in the broad main discussion..
\end{abstract}

\section{Introduction}
Our work \cite{BFC2025_Dir_ISAC} introduces an integrated OTFS-RSMA framework for Integrated Sensing and Communication (ISAC) in Low Earth Orbit (LEO) satellite systems. Accurate performance analysis and resource allocation design require precise models for sensing accuracy (via CRB) and communication reliability (via SINR). This supplement details the derivation of these key metrics, considering practical impairments such as variable echo gain, imperfect CSI (ICSI), and imperfect SIC (ISIC). The inclusion of these impairments follows methodologies validated in relevant literature for sensing and communication systems. 

\section{System Model Overview}

We consider the downlink of an ISAC system where a satellite serves $K$ users using RSMA over an OTFS modulation scheme. Key variables represented in the Delay-Doppler (DD) domain are:

\begin{itemize}
    \item $\mathbf{X}_{\mathrm{DD}} \in \mathbb{C}^{N_{dd} \times 1}$: Transmitted DD signal vector, $N_{dd}=MN$. $\mathbf{X}_{\mathrm{DD}} = \boldsymbol{\mathcal{P}}_c s_c + \sum_{j=1}^K \boldsymbol{\mathcal{P}}_{p,j} s_{p,j}$.
    \item $\boldsymbol{\mathcal{P}}_c, \boldsymbol{\mathcal{P}}_{p,j} \in \mathbb{C}^{N_{dd} \times 1}$: Precoding vectors for common and private streams. $P_{\text{tot}} = \|\boldsymbol{\mathcal{P}}_c\|^2 + \sum_{k=1}^K \|\boldsymbol{\mathcal{P}}_{p,k}\|^2$.
    \item $\mathbf{H}_k \in \mathbb{C}^{N_{dd} \times N_{dd}}$: True effective DD channel matrix for user $k$.
    \item $\mathbf{N}_k \in \mathbb{C}^{N_{dd} \times 1}$: AWGN vector, $\mathbf{N}_k \sim \mathcal{CN}(\mathbf{0}, \sigma_n^2 \mathbf{I})$.
    \item $\mathbf{Y}_k \in \mathbb{C}^{N_{dd} \times 1}$: Received DD signal vector, $\mathbf{Y}_k = \mathbf{H}_k \mathbf{X}_{\mathrm{DD}} + \mathbf{N}_k$.
    \item $\mathbf{E}_k \in \mathbb{C}^{N_{dd} \times N_{dd}}$: Channel estimation error matrix, elements i.i.d. $\mathcal{CN}(0, \sigma_e^2)$.
    \item $\hat{\mathbf{H}}_k \in \mathbb{C}^{N_{dd} \times N_{dd}}$: Estimated DD channel matrix, $\hat{\mathbf{H}}_k = \mathbf{H}_k + \mathbf{E}_k$.
    \item $\mathbf{w}_{c,k}, \mathbf{w}_{p,k} \in \mathbb{C}^{1 \times N_{dd}}$: LMMSE filter row vectors (or $N_{dd} \times 1$ column vectors, adjusting subsequent math) designed using $\hat{\mathbf{H}}_k$.
    \item $\Theta_k \in [0, 1]$: ISIC imperfection factor for user $k$    
\end{itemize}

For sensing, we consider a monostatic scenario targeting parameters delay ($\tau_T$) and Doppler ($\nu_T$). The echo model depends on the specific assumptions (e.g., variable gain).

\section{Derivation of Cramér-Rao Bounds (CRB) with Variable Gain}
\label{sec:crb_derivation}

This section details the derivation of the Fisher Information Matrix (FIM) for estimating delay ($\tau_T$) and Doppler ($\nu_T$) when the complex echo path gain $\alpha$ is considered variable with delay, specifically $\alpha = \alpha(\tau_T)$, based on the model presented in the main paper. The noise variance $\sigma_{\text{echo}}^2$ is denoted as $\sigma^2$ herein for brevity. The fundamental principles of FIM calculation for Gaussian noise models are well-established \cite{Kay1993}. 

\subsection{Step 1: Expressing the Mean Received DD Signal \texorpdfstring{$\mu[l,k]$}{mu[l,k]}}

Our starting point is the signal received in the Time-Frequency (TF) domain, $M[n,i]$, incorporating the transmitted TF symbol $X[n,i]$, the variable gain $\alpha(\tau_T)$, the target reflectivity $\beta_T$, and the standard phase shifts due to Doppler $\nu_T$ and delay $\tau_T$:

\begin{equation}
    M[n,i] \approx \alpha(\tau_T) \beta_T X[n,i] e^{j2\pi (\nu_T n T - \tau_T i \Delta f)}.
    \label{eq:supp_M_ni_variable_alpha}
\end{equation}

Here, $n=0, \dots, N-1$, $i=0, \dots, M-1$, $T$ is the OTFS symbol duration in the TF domain, and $\Delta f$ is the subcarrier spacing. The inclusion of a parameter-dependent gain $\alpha(\tau_T)$ follows methodological precedents in sensing literature where signal amplitude depends on estimated parameters, such as in extended target analysis or parameter-dependent reflectivity models \cite{Boyd2020}. The mean received signal in the Delay-Doppler (DD) domain, $\mu[l,k]$, obtained via the Symplectic Finite Fourier Transform (SFFT) \cite{Hadani2017}, is: 

\begin{equation}
    \mu[l,k] = \frac{1}{\sqrt{MN}} \sum_{n=0}^{N-1} \sum_{i=0}^{M-1} M[n,i] e^{-j2\pi \left(\frac{nl}{N} - \frac{ik}{M}\right)}.
    \label{eq:supp_sfft_def_mu}
\end{equation}

Substituting \eqref{eq:supp_M_ni_variable_alpha} into \eqref{eq:supp_sfft_def_mu} yields:

\begin{equation}
    \mu[l,k] = \frac{\alpha(\tau_T) \beta_T}{\sqrt{MN}} \sum_{n=0}^{N-1} \sum_{i=0}^{M-1} X[n,i] \exp\left[ j2\pi \left( n( \nu_T T - \frac{l}{N}) - i(\tau_T \Delta f - \frac{k}{M}) \right) \right].
    \label{eq:supp_mu_lk_final_step1}
\end{equation}

This expression forms the basis for calculating the FIM elements.

\subsection{Step 2: Calculating the Derivatives \texorpdfstring{$\partial \mu[l,k] / \partial \nu_T$}{dmu/dnu} and \texorpdfstring{$\partial \mu[l,k] / \partial \tau_T$}{dmu/dtau}}

We differentiate \eqref{eq:supp_mu_lk_final_step1} with respect to $\nu_T$ and $\tau_T$.

\subsubsection{Derivative with respect to Doppler ($\nu_T$)}

Differentiating with respect to $\nu_T$ affects only the phase term related to $n$:

\begin{equation}
\frac{\pr \mu[l,k]}{\pr \nu_T} = \frac{j2\pi T \alpha(\tau_T) \beta_T}{\sqrt{MN}} \sum_{n=0}^{N-1} \sum_{i=0}^{M-1} (n X[n,i]) \exp\left[ j2\pi \left( n( \nu_T T - \frac{l}{N}) - i(\tau_T \Delta f - \frac{k}{M}) \right) \right].
\label{eq:supp_dmu_dnu_exact}
\end{equation}

Let $\mathbf{d}_{\nu}$ denote the vector containing these elements for all $(l,k)$.

\subsubsection{Derivative with respect to Delay ($\tau_T$)}

Here, both $\alpha(\tau_T)$ and the phase term related to $i$ depend on $\tau_T$. Applying the product rule and using the specific gain model $\pr \alpha / \pr \tau_T = -2\alpha(\tau_T) / \tau_T$, we obtain:

\begin{equation}
\small
\frac{\pr \mu[l,k]}{\pr \tau_T} = \left( \frac{\pr \alpha(\tau_T)}{\pr \tau_T} \right) (\dots) + \alpha(\tau_T) \frac{\pr}{\pr \tau_T}(\dots) = \frac{-2}{\tau_T} \mu[l,k] - \frac{j2\pi \Delta f \alpha(\tau_T) \beta_T}{\sqrt{MN}} \sum_{n=0}^{N-1} \sum_{i=0}^{M-1} (i X[n,i]) \exp\left[ j2\pi \left( n(\dots) - i(\dots) \right) \right].
\label{eq:supp_dmu_dtau_exact}
\end{equation}

Let $\mathbf{d}_{\tau}$ denote the vector containing these elements. We can identify $\mathbf{d}_{\tau} = \mathbf{d}_{\text{gain}} + \mathbf{d}_{\text{phase},\tau}$, where $\mathbf{d}_{\text{gain}} = \frac{-2}{\tau_T}\boldsymbol{\mu}$ and $\mathbf{d}_{\text{phase},\tau}$ is the second term in \eqref{eq:supp_dmu_dtau_exact}. The application of the chain rule when differentiating with respect to $\tau_T$, accounting for both phase and amplitude dependence on the parameter, is a standard step in FIM calculations with such models \cite{Boyd2020}. 

\subsection{Step 3: Calculating the FIM Elements \texorpdfstring{$I_{ij}^{(\text{exact})}$}{Iij\_exact}}

The FIM elements are calculated using the standard formula for complex Gaussian observations \cite{Kay1993}: 

\begin{equation}
    [\mathbf{I}^{(\text{exact})}]_{i,j} = \frac{2}{\sigma^2} \text{Re}\{ (\mathbf{d}_{i}^{(\text{exact})})^{\herm} \mathbf{d}_{j}^{(\text{exact})} \}.
\end{equation}

The structure of these calculations involving derivatives in the DD domain aligns with FIM/CRB analyses specific to OTFS sensing \cite{Wu2024}. 

\subsubsection{Element \texorpdfstring{$I_{\nu\nu}^{(\text{exact})}$}{I\_nunu\_exact}}

\begin{equation}
I_{\nu\nu}^{(\text{exact})} = \frac{2}{\sigma^2} \| \mathbf{d}_{\nu} \|^2 = \frac{2 (2\pi T)^2 |\alpha(\tau_T)|^2 |\beta_T|^2}{MN \sigma^2} \sum_{l,k} \left| \sum_{n,i} n X[n,i] e^{j\phi_{n,i,l,k}} \right|^2,
\label{eq:supp_Inunu_exact_final}
\end{equation}

where $\phi_{n,i,l,k}$ is the phase term from \eqref{eq:supp_mu_lk_final_step1}.

\subsubsection{Element \texorpdfstring{$I_{\tau\tau}^{(\text{exact})}$}{I\_tautau\_exact}}

Using $\mathbf{d}_{\tau} = \mathbf{d}_{\text{gain}} + \mathbf{d}_{\text{phase},\tau}$:

\begin{equation}
I_{\tau\tau}^{(\text{exact})} = \frac{2}{\sigma^2} \left( \|\mathbf{d}_{\text{gain}}\|^2 + \|\mathbf{d}_{\text{phase},\tau}\|^2 + 2 \text{Re}\{ \mathbf{d}_{\text{gain}}^{\herm} \mathbf{d}_{\text{phase},\tau} \} \right).
\label{eq:supp_Itautau_exact_components}
\end{equation}

The individual components are:

\begin{itemize}
\item $\|\mathbf{d}_{\text{gain}}\|^2 = \frac{4}{\tau_T^2} \|\boldsymbol{\mu}\|^2$.
\item $\|\mathbf{d}_{\text{phase},\tau}\|^2 = \frac{(2\pi \Delta f)^2 |\alpha(\tau_T)|^2 |\beta_T|^2}{MN} \sum_{l,k} \left| \sum_{n,i} i X[n,i] e^{j\phi_{n,i,l,k}} \right|^2$.
\item $2 \text{Re}\{ \mathbf{d}_{\text{gain}}^{\herm} \mathbf{d}_{\text{phase},\tau} \} = -\frac{4}{\tau_T} \text{Re}\{ \boldsymbol{\mu}^{\herm} \mathbf{d}_{\text{phase},\tau} \}$.
\end{itemize}

Substituting these gives the full expression for $I_{\tau\tau}^{(\text{exact})}$.

\subsubsection{Element \texorpdfstring{$I_{\tau\nu}^{(\text{exact})}$}{I\_taunu\_exact} (Off-Diagonal)}

Using $\mathbf{d}_{\tau} = \mathbf{d}_{\text{gain}} + \mathbf{d}_{\text{phase},\tau}$ and $\mathbf{d}_{\nu}$:

\begin{equation}
I_{\tau\nu}^{(\text{exact})} = \frac{2}{\sigma^2} \left( \text{Re}\{ \mathbf{d}_{\text{gain}}^{\herm} \mathbf{d}_{\nu} \} + \text{Re}\{ \mathbf{d}_{\text{phase},\tau}^{\herm} \mathbf{d}_{\nu} \} \right).
\label{eq:supp_Itaunu_exact_components}
\end{equation}

The individual components involve inner products between the vectors derived in Step 2. Specifically:

\begin{itemize}
\item $\text{Re}\{ \mathbf{d}_{\text{gain}}^{\herm} \mathbf{d}_{\nu} \} = -\frac{2}{\tau_T} \text{Re}\{ \boldsymbol{\mu}^{\herm} \mathbf{d}_{\nu} \}$.
\item $\text{Re}\{ \mathbf{d}_{\text{phase},\tau}^{\herm} \mathbf{d}_{\nu} \}$ is the real part of the inner product between the phase derivative w.r.t. $\tau_T$ and the derivative w.r.t. $\nu_T$.
\end{itemize}

\subsection{Step 4: Inverting the FIM for the Exact CRB}

The CRB matrix is $\mathbf{CRB}^{(\text{exact})} = (\mathbf{I}^{(\text{exact})})^{-1}$. For the $2 \times 2$ case:

\begin{equation}
\mathbf{CRB}^{(\text{exact})} = \frac{1}{\det(\mathbf{I}^{(\text{exact})})} \begin{pmatrix} I_{\nu\nu}^{(\text{exact})} & -I_{\tau\nu}^{(\text{exact})} \\ -I_{\tau\nu}^{(\text{exact})} & I_{\tau\tau}^{(\text{exact})} \end{pmatrix},
\label{eq:supp_CRB_matrix_exact_symbolic}
\end{equation}

where $\det(\mathbf{I}^{(\text{exact})}) = I_{\tau\tau}^{(\text{exact})} I_{\nu\nu}^{(\text{exact})} - (I_{\tau\nu}^{(\text{exact})})^2$.

The specific bounds are the diagonal elements:

\begin{equation}
\boxed{\mathrm{CRB}(\tau_T)^{(\text{exact})} = \frac{I_{\nu\nu}^{(\text{exact})}}{\det(\mathbf{I}^{(\text{exact})})}}
\label{eq:supp_CRB_tau_exact_final}
\end{equation}

\begin{equation}
\boxed{\mathrm{CRB}(\nu_T)^{(\text{exact})} = \frac{I_{\tau\tau}^{(\text{exact})}}{\det(\mathbf{I}^{(\text{exact})})}}
\label{eq:supp_CRB_nu_exact_final}
\end{equation}

Substituting the full expressions for the FIM elements derived in Step 3 yields the exact CRBs under the variable gain model. While the final algebraic form is complex, we present it here for completeness.

First, let us define some intermediate quantities based on the derivative vectors $\mathbf{d}_{\nu}$ (from \eqref{eq:supp_dmu_dnu_exact}), $\mathbf{d}_{\text{phase},\tau}$ (the second term in \eqref{eq:supp_dmu_dtau_exact}), and the mean signal vector $\boldsymbol{\mu}$ (whose elements are $\mu[l,k]$ from \eqref{eq:supp_mu_lk_final_step1}):
\begin{equation}
S_n \triangleq \sum_{l=0}^{M-1} \sum_{k=0}^{N-1} \left| \sum_{n=0}^{N-1} \sum_{i=0}^{M-1} n X[n,i] e^{j\phi_{n,i,l,k}} \right|^2
\end{equation}
\begin{equation}
S_i \triangleq \sum_{l=0}^{M-1} \sum_{k=0}^{N-1} \left| \sum_{n=0}^{N-1} \sum_{i=0}^{M-1} i X[n,i] e^{j\phi_{n,i,l,k}} \right|^2
\end{equation}
\begin{equation}
C_{\tau\nu} \triangleq \text{Re}\{ \mathbf{d}_{\text{phase},\tau}^{\herm} \mathbf{d}_{\nu} \}
\end{equation}

\begin{equation}
C_{\mu\tau} \triangleq \text{Re}\{ \boldsymbol{\mu}^{\herm} \mathbf{d}_{\text{phase},\tau} \}
\end{equation}

\begin{equation}
C_{\mu\nu} \triangleq \text{Re}\{ \boldsymbol{\mu}^{\herm} \mathbf{d}_{\nu} \}
\end{equation}

\begin{equation}
P_{\mu} \triangleq \|\boldsymbol{\mu}\|^2 = \sum_{l=0}^{M-1} \sum_{k=0}^{N-1} |\mu[l,k]|^2
\end{equation}

where $\phi_{n,i,l,k} = 2\pi [ n( \nu_T T - l/N) - i(\tau_T \Delta f - k/M) ]$, and $\sigma^2 = \sigma_{\text{echo}}^2$. Also let $|\alpha|^2 = |\alpha(\tau_T)|^2$ and $|\beta|^2 = |\beta_T|^2$.

Using these definitions, the FIM elements are:

\begin{align}
I_{\nu\nu}^{(\text{exact})} &= \frac{2 (2\pi T)^2 |\alpha|^2 |\beta|^2}{MN \sigma^2} S_n \label{eq:supp_Inunu_expanded_intermediate}\\
I_{\tau\tau}^{(\text{exact})} &= \frac{8 P_{\mu}}{\sigma^2 \tau_T^2} + \frac{2(2\pi \Delta f)^2 |\alpha|^2 |\beta|^2}{MN \sigma^2} S_i - \frac{8}{\sigma^2 \tau_T} C_{\mu\tau} \label{eq:supp_Itautau_expanded_intermediate}\\
I_{\tau\nu}^{(\text{exact})} &= \frac{2}{\sigma^2} C_{\tau\nu} - \frac{4}{\sigma^2 \tau_T} C_{\mu\nu} \label{eq:supp_Itaunu_expanded_intermediate}
\end{align}

The determinant of the FIM is $\det(\mathbf{I}^{(\text{exact})}) = I_{\tau\tau}^{(\text{exact})} I_{\nu\nu}^{(\text{exact})} - (I_{\tau\nu}^{(\text{exact})})^2$. Substituting the expressions above:

\begin{align}
\det(\mathbf{I}^{(\text{exact})}) = &\left( \frac{8 P_{\mu}}{\sigma^2 \tau_T^2} + \frac{2(2\pi \Delta f)^2 |\alpha|^2 |\beta|^2}{MN \sigma^2} S_i - \frac{8 C_{\mu\tau}}{\sigma^2 \tau_T} \right) \left( \frac{2 (2\pi T)^2 |\alpha|^2 |\beta|^2}{MN \sigma^2} S_n \right) - \left( \frac{2 C_{\tau\nu}}{\sigma^2} - \frac{4 C_{\mu\nu}}{\sigma^2 \tau_T} \right)^2 \label{eq:supp_Det_expanded_intermediate}
\end{align}

Now, we substitute these into the CRB formulas \eqref{eq:supp_CRB_tau_exact_final} and \eqref{eq:supp_CRB_nu_exact_final}.

The explicit expression for $\mathrm{CRB}(\tau_T)^{(\text{exact})}$ is:

\begin{empheq}[box=\fbox]{equation} \label{eq:supp_CRB_tau_explicit_final}
\mathrm{CRB}(\tau_T)^{(\text{exact})} = \frac{\frac{2 (2\pi T)^2 |\alpha(\tau_T)|^2 |\beta_T|^2}{MN \sigma_{\text{echo}}^2} S_n}{\left( \frac{8 P_{\mu}}{\sigma_{\text{echo}}^2 \tau_T^2} + \frac{2(2\pi \Delta f)^2 |\alpha(\tau_T)|^2 |\beta_T|^2}{MN \sigma_{\text{echo}}^2} S_i - \frac{8 C_{\mu\tau}}{\sigma_{\text{echo}}^2 \tau_T} \right) \left( \frac{2 (2\pi T)^2 |\alpha(\tau_T)|^2 |\beta_T|^2}{MN \sigma_{\text{echo}}^2} S_n \right) - \left( \frac{2 C_{\tau\nu}}{\sigma_{\text{echo}}^2} - \frac{4 C_{\mu\nu}}{\sigma_{\text{echo}}^2 \tau_T} \right)^2}
\end{empheq}

The explicit expression for $\mathrm{CRB}(\nu_T)^{(\text{exact})}$ is:

\begin{empheq}[box=\fbox]{equation} \label{eq:supp_CRB_nu_explicit_final}
\mathrm{CRB}(\nu_T)^{(\text{exact})} = \frac{\frac{8 P_{\mu}}{\sigma_{\text{echo}}^2 \tau_T^2} + \frac{2(2\pi \Delta f)^2 |\alpha(\tau_T)|^2 |\beta_T|^2}{MN \sigma_{\text{echo}}^2} S_i - \frac{8 C_{\mu\tau}}{\sigma_{\text{echo}}^2 \tau_T}}{\left( \frac{8 P_{\mu}}{\sigma_{\text{echo}}^2 \tau_T^2} + \frac{2(2\pi \Delta f)^2 |\alpha(\tau_T)|^2 |\beta_T|^2}{MN \sigma_{\text{echo}}^2} S_i - \frac{8 C_{\mu\tau}}{\sigma_{\text{echo}}^2 \tau_T} \right) \left( \frac{2 (2\pi T)^2 |\alpha(\tau_T)|^2 |\beta_T|^2}{MN \sigma_{\text{echo}}^2} S_n \right) - \left( \frac{2 C_{\tau\nu}}{\sigma_{\text{echo}}^2} - \frac{4 C_{\mu\nu}}{\sigma_{\text{echo}}^2 \tau_T} \right)^2}
\end{empheq}

These final CRB expressions \eqref{eq:supp_CRB_tau_explicit_final} and \eqref{eq:supp_CRB_nu_explicit_final} provide the theoretical lower bounds on estimation variance under the specified model incorporating variable gain.

\section{Refined SINR Analysis under Imperfect CSI and SIC}
\label{sec:supp_refined_sinr}

The performance evaluation of the proposed OTFS-RSMA framework fundamentally relies on accurate SINR expressions. We refine the baseline SINR by explicitly incorporating the effects of imperfect CSI (ICSI) and imperfect SIC (ISIC), drawing upon established modeling techniques from the RSMA and OTFS literature.

We maintain the system model where the received DD signal is $\mathbf{Y}_k = \mathbf{H}_k \mathbf{X}_{\mathrm{DD}} + \mathbf{N}_k$. The estimated channel is $\hat{\mathbf{H}}_k = \mathbf{H}_k + \mathbf{E}_k$, with $\mathbf{E}_k$ being the error matrix. LMMSE filters $\mathbf{w}_{c,k}, \mathbf{w}_{p,k}$ are designed using $\hat{\mathbf{H}}_k$.

\subsection{Modeling Imperfections}

\subsubsection{Imperfect SIC (ISIC)}
Residual interference from imperfect cancellation of the common stream $s_c$ is modeled using an imperfection factor $\Theta_k \in [0, 1]$. The residual interference power affecting the private stream decoding is approximated as $P_{I,c}^{\text{res}} = \Theta_k |\mathbf{w}_{p,k} \hat{\mathbf{H}}_k \boldsymbol{\mathcal{P}}_c|^2$. This modeling approach, using a scaling factor for residual power, is consistent with analyses of ISIC in RSMA systems \cite{Karim2025}. 

\subsubsection{Imperfect CSI (ICSI) Impact on LMMSE Filtering}

The additive error model $\hat{\mathbf{H}}_k = \mathbf{H}_k + \mathbf{E}_k$, where error elements have variance $\sigma_e^2$, is a standard approach for ICSI \cite{Daesung2022}. The mismatch between the true channel $\mathbf{H}_k$ and the estimate $\hat{\mathbf{H}}_k$ used for LMMSE filter design ($\mathbf{w}_k$) results in suboptimal filtering and introduces additional distortion terms. The total variance due to ICSI across all streams, after filtering, is often approximated as an effective noise enhancement term proportional to the error variance $\sigma_e^2$, the filter norm $\|\mathbf{w}_k\|^2$, and the total transmitted power $P_{\text{tot}}$, i.e., $\sigma_e^2 \|\mathbf{w}_k\|^2 P_{\text{tot}}$ \cite{Daesung2022, Singh2022, Mishra2022}.

This approximation accounts for the increased noise and residual interference due to filter mismatch.

\subsection{Derivation of Refined SINR Expressions}
We derive the SINR considering the desired signal power based on $\hat{\mathbf{H}}_k$ (consistent with filter design) and incorporating interference and noise terms accounting for ISIC and ICSI.

\subsubsection{Common Stream SINR ($\mathrm{SINR}_{c,k}$)}

The desired signal power is $P_{S,c} = |\mathbf{w}_{c,k} \hat{\mathbf{H}}_k \boldsymbol{\mathcal{P}}_c|^2$. The denominator includes interference from private streams based on $\hat{\mathbf{H}}_k$ and the effective noise term capturing AWGN and ICSI effects:

\begin{equation}
    D_c \approx \sum_{j=1}^K |\mathbf{w}_{c,k} \hat{\mathbf{H}}_k \boldsymbol{\mathcal{P}}_{p,j}|^2 + \|\mathbf{w}_{c,k}\|^2 (\sigma_n^2 + \sigma_e^2 P_{\text{tot}}).
\end{equation}

The refined SINR is thus:

\begin{empheq}[box=\fbox]{equation}
\label{eq:supp_sinr_c_final_boxed}
\mathrm{SINR}_{c,k}^{\text{(ref)}} \approx \frac{| \mathbf{w}_{c,k} \hat{\mathbf{H}}_k \boldsymbol{\mathcal{P}}_c |^2 }{D_c}
\end{empheq}

\subsubsection{Private Stream SINR ($\mathrm{SINR}_{p,k}$)}

The desired signal power is $P_{S,p} = |\mathbf{w}_{p,k} \hat{\mathbf{H}}_k \boldsymbol{\mathcal{P}}_{p,k}|^2$. The denominator includes interference from other private streams, residual common stream interference (scaled by $\Theta_k$), and the effective noise term:

\begin{equation}
    D_p \approx \sum_{j \neq k} |\mathbf{w}_{p,k} \hat{\mathbf{H}}_k \boldsymbol{\mathcal{P}}_{p,j}|^2 + \Theta_k |\mathbf{w}_{p,k} \hat{\mathbf{H}}_k \boldsymbol{\mathcal{P}}_c|^2 + \|\mathbf{w}_{p,k}\|^2 (\sigma_n^2 + \sigma_e^2 P_{\text{tot}}).
\end{equation}

The refined SINR, explicitly accounting for ISIC via $\Theta_k$ and ICSI via the effective noise term, is:

\begin{empheq}[box=\fbox]{equation}
\label{eq:supp_sinr_p_final_boxed}
\mathrm{SINR}_{p,k}^{\text{(ref)}} \approx \frac{| \mathbf{w}_{p,k} \hat{\mathbf{H}}_k \boldsymbol{\mathcal{P}}_{p,k} |^2 }{D_p}
\end{empheq}

\vspace{1mm} 
\bibliographystyle{IEEEtran}
\bibliography{refs}

\end{document}